# Quantitative Histopathology of Stained Tissues using Color Spatial Light Interference Microscopy (cSLIM)

*Hassaan Majeed[1], Adib Keikhosravi[2], Mikhail E. Kandel[3], Tan H. Nguyen[4], Yuming Liu[2], Andre Kajdacsy-Balla[5], Krishnarao Tangella[6], Kevin W. Eliceiri[2], Gabriel Popescu[1,3,*].*

[1] Quantitative Light Imaging (QLI) Lab, Department of Bioengineering, Beckman Institute of Advanced Science and Technology, University of Illinois at Urbana Champaign, 405 N. Matthews, Urbana, IL 61801, USA.

[2] Laboratory for Optical and Computational Instrumentation (LOCI), Department of Biomedical Engineering, University of Wisconsin-Madison, Madison, WI 53706, USA.

[3] Quantitative Light Imaging (QLI) Lab, Department of Electrical and Computer Engineering, Beckman Institute of Advanced Science and Technology, University of Illinois at Urbana Champaign, 405 N. Matthews, Urbana, IL 61801, USA.

[4] Biomedical Optics and Biophotonics group, Department of Electrical Engineering and Computer Science, Massachusetts Institute of Technology, Cambridge, MA 02142, USA.

[5] Department of Pathology, University of Illinois at Chicago, 840 South Wood Street, Suite 130 CSN, Chicago, IL 60612, USA.

[6] Christie Clinic, 1400 West Park Street, Urbana, IL 61801, USA.

*Abstract:* Tissue biopsy evaluation in the clinic is in need of quantitative disease markers for diagnosis and, most importantly, prognosis. Among the new technologies, quantitative phase imaging (QPI) has demonstrated promise for histopathology because it reveals intrinsic tissue nanoarchitecture through the refractive index. However, a vast majority of past QPI investigations have relied on imaging unstained tissues, which disrupts the established specimen processing. Here we present color spatial light interference microscopy (cSLIM) as a new whole slide imaging modality that performs interferometric imaging with a color detector array. As a result, cSLIM yields in a single scan both the intrinsic tissue phase map and the standard color bright-field image, familiar to the pathologist. Our results on 196 breast cancer patients indicate that cSLIM can provide not only diagnostic but also prognostic information from the alignment of collagen fibers in the tumor microenvironment. The effects of staining on the tissue phase maps were corrected by a simple mathematical normalization. These characteristics are likely to reduce barriers to clinical translation for the new cSLIM technology.

**1. Introduction**

The field of histopathology is undergoing significant transformation, thanks to advances in imaging and computation technology. For example, in 2017 Philips received FDA approval for digital pathology in primary diagnosis [1-3], while the government of Kuwait approved Leica Biosystems digital pathology as its standard of care [4]. Computer assisted image analysis (CAIA) of whole slide images (WSI) is slated to play an important role in corroborating information across multiple laboratories and, ultimately, improving diagnosis accuracy and patient outcome. The current gold standard in histopathology involves the microscopic investigation of a surgical specimen by a trained pathologist. The tissue is processed in the form of a thin section placed on a cover glass. Because such tissue sections are completely transparent under visible light, standard protocols include a staining procedure. As such, the pathologist is not observing the tissue morphology itself, but rather, a stain distribution that correlates with morphology. The contrast and color balance varies not only with the pathologist preference [see, e.g., Figure 1 in Ref. [5]] but also with the specimen processing, staining procedure and illumination. Staining variability has been identified as a major obstacle in producing consistent results across multiple specimens when using machine-learning algorithms on stained tissues [6-8]. To take advantage of the digital data that are increasingly available from commercial WSI, numerical solutions have been proposed to correct the staining inconsistency in post-processing [9-11].

Label-free microscopy techniques are promising for meeting this current challenge in pathology. By generating intrinsic contrast in images based on physical properties of the specimen, these methods minimize inter-observer variation and subjectivity in the evaluation. Their label-free nature also simplifies the process of designing supervised learning schemes for automated image analysis as consistent feature values are expected between training and deployment [12-14]. Furthermore, by relying on novel contrast mechanisms yet unfamiliar to traditional pathology, these methods can also elucidate new cancer biology and introduce new markers of prognosis [15-17]. A number of label-free methods have been proposed for histopathology in the past. These can be broadly grouped into two categories: vibrational spectroscopy methods [including Fourier Transform Infrared spectroscopy (FT-IR) [14,18,19] and Raman spectroscopy (RS) [20-22]] and non-linear optical microscopy methods [including second harmonic generation microscopy (SHGM) [15,23], endogenous two photon excited fluorescence (TPEF) [24] and third harmonic generation microscopy (THGM) [25]]. Each technique in these groups brings with it unique features in terms of chemical specificity and physical information extracted. However, these methods also face specific challenges for clinical translation, as they can pose new requirements on sample preparation and optical instrumentation. For example, FT-IR yields lower spatial resolution and many orders of magnitude lower throughput than conventional bright-field microscopy. RS is also much slower than conventional microscopy. In addition, these techniques have been applied mainly on unstained tissue sections. Since staining is an inevitable part of standard tissue assessment, this complicates the process of clinical translation for these techniques. While some non-linear optical microscopy studies have been performed on stained tissues[13,15,23], techniques in this group also have much lower throughput than conventional microscopy, require expensive ultrashort-pulsed lasers and are only able to generate contrast in certain structures (such as collagen fibers, for example, in SHGM)[26]. All these methods require special illumination and detection systems that are difficult to combine with bright-field microscopes currently present in pathology labs.

Quantitative phase imaging (QPI)[27], is an emerging technology that yields intrinsic contrast by mapping the optical pathlength difference (OPD) across the tissue specimen. QPI has been successfully employed for a number of quantitative histopathology investigations[28], including diagnosis of prostate cancer[29], prediction of recurrence in patients after prostatectomy[16,30], colon cancer screening[31], breast cancer diagnosis[32], Gleason grading of prostate cancer[12], collagen fiber orientation measurement in breast tissue[33], assessment of malignancy in bile ducts[34] and early prediction of risk of colon cancer[35]. QPI systems can be deployed as add-on modules to conventional microscopes and provide similar resolution and throughput to those in traditional systems[31,36]. Thus, they have the potential to minimally disrupt the current clinical pipeline. However, almost all of the work done thus far using QPI has involved unstained tissue biopsies. These QPI methods would require a pathologist to prepare an additional unstained tissue section which needs to be scanned separately if both traditional and new markers for disease are to be obtained.

Here, we propose a technical solution for extracting *intrinsic* tissue morphology information from *stained* tissues sections, yet, unaffected by variability in tissue staining. We developed color spatial light interference microscopy (cSLIM), in which the image is overlapped with a reference wave to reveal a detailed map of the OPD across the tissue. The cSLIM image is produced by the refractive index of the tissue, which is an intrinsic morphological marker correlated with disease [29]. The cSLIM instrument is an advanced form of QPI. However, unlike previous QPI methods applied for diagnosis[16,27,29-31,33,36], cSLIM operates on existing *stained* tissue slides and, in a single scan, produces both a bright-field map and a phase map for intrinsic contrast. Using specimens prepared under the standard protocol, cSLIM yields simultaneously the typical image that the pathologist is accustomed to (e.g., H&E, immunochemical stains, etc.) and a quantitative phase image, which provides new information, currently not available in bright-field images (e.g., collagen fiber orientation[33]). Importantly, the effects of the stain on the phase image can be removed completely by using a simple mathematical operation. The cSLIM throughput as a WSI scanner is comparable with that of commercial (bright-field) instruments [31]. We anticipate that computer algorithms (see, e.g.,[8,37-39]) currently used on stained tissue images will translate, upon adjustments, with strong performance on our quantitative images that are free of stain variability.

In order to illustrate the potential of the new cSLIM technology for histopathology applications, we demonstrate its value for breast cancer diagnosis and prognosis. Breast cancer is a significant global health problem, being the second most commonly diagnosed cancer worldwide according to the latest World Health Organization (WHO) statistics [40]. Within the US, while rates of mortality have been consistently falling, rates of incidence have been on the rise with 266,120 new cases of invasive breast cancer expected in 2018 [41]. This indicates that an increasing number of new diagnostic and prognostic evaluations are being performed on patients by pathology labs nationwide. Robust methods for quantitating disease signatures in breast tissue, independent of stain variation, remain highly desirable. Furthermore, breast cancer is a heterogeneous disease. Due to variations in tumor type and, therefore, patient responses to different therapies, the current set of prognostic indicators do not provide sufficient information for all patients [15]. There is, thus, a need for novel biomarkers that inform objectively on disease aggressiveness to account for individual variation [42,43]. QPI-based biomarkers can be significant in this regard since they can be easily integrated into the current workflow and, with the advent of computational pathology, can be extracted rapidly and reproducibly [13,16]. We show that cSLIM can separate benign and malignant lesions in stain-independent phase images of breast tissue. Furthermore, we demonstrate that cSLIM reveals the alignment and orientation of tumor adjacent collagen fibers, a histological marker with known prognostic value for breast cancer patients [15]. We classified patients based on the cSLIM-measured value of this marker and showed its predictive power though survival analysis.

This paper is organized as follows. We first introduce the cSLIM instrument and describe the optical problem of interferometry with an RGB camera. Next, we describe the normalization process that eliminates the stain dependent signal from our cSLIM images. Finally, we demonstrate the ability of our stained tissue analysis system to diagnose malignancy in tissue and to detect aligned collagen fiber-based prognostic markers.

## 2. Results and Discussion

*a) cSLIM optical setup*

The cSLIM optical setup is illustrated in Fig. 1. The optical train builds on the principle of spatial light interference microscopy (SLIM) [44]. The SLIM module (CellVista SLIM Pro, Phi Optics, Inc.) is placed at the output port of a commercial microscope. The conjugate image plane at the microscope output port is imaged onto the camera using a 4f system formed by lenses $L_1$ and $L_2$. At the Fourier plane of lens $L_1$ a spatial light modulator (SLM, Boulder Nonlinear Systems) is used to modulate the phase difference between the scattered and incident light. As illustrated in Figs. 1 (c) and (d), four phase shifts are employed, $\varphi = 0, \pi/2, \pi, 3\pi/2$ rad, and intensity images are recorded by the camera for each shift. Combining the four intensity images [44], we obtain the quantitative phase image, free of amplitude effects.

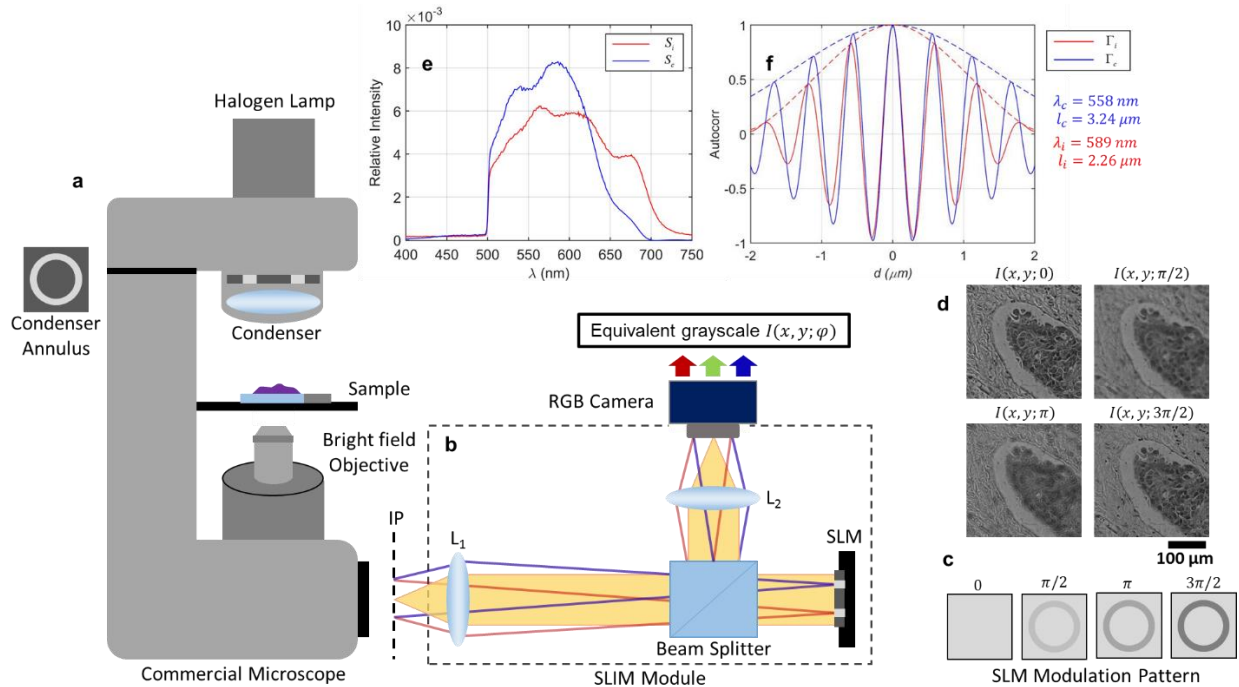

**Figure 1.** cSLIM optical setup. **(a)** A commercial microscope with phase contrast illumination and bright-field objective. **(b)** SLIM module placed at the microscope output port, equipped with color RGB camera. **(c)** Phase shifts imparted between scattered and incident light by the SLM in the Fourier plane of $L_1$. **(d)** Four intensity frames obtained for each phase modulation by weighted sum of the red, green and blue channel images of the RGB camera **(e)** Spectra and **(f)** Temporal autocorrelation (real part) for RGB and grayscale imaging. IP, Image Plane, SLM, Spatial light Modulator.

There are several significant modifications in the current cSLIM system with respect to previous reports [44], as follows. First, in place of the phase contrast microscope objective used in SLIM we employ a 40x/0.75 NA bright-field objective. The use of a bright-field objective allows us to obtain typical bright-field images, i.e., H&E stained histology images that are the mainstay of histopathology in the clinic. The annular condenser ring is retained from the phase contrast microscope. Second, the grayscale camera is replaced by an RGB camera (Carl Zeiss Axiocam MRc) which provides red, green, and blue channels. Finally, because we use an RGB camera, the detected phase is that of the sum of the autocorrelation functions for each spectral channel (see below). As a result, the calibration of the SLM and the physical meaning of the measured phase are different from the previous reports.

For each phase shift, we acquire three intensity frames corresponding to the red, green and blue channels of the camera: $R(x,y;\varphi)$, $G(x,y;\varphi)$ and $B(x,y;\varphi)$, respectively. In each case, these channels are combined linearly to obtain an equivalent grayscale intensity image, as

$$I(x,y;\varphi) = rR(x,y;\varphi) + gG(x,y;\varphi) + bB(x,y;\varphi), \qquad (1)$$

where $r, g, b$ are constants such that $r+g+b=1$. Throughout the manuscript, we use $r=0.1$, $g=0.6$, $b=0.3$. This weighting was determined empirically, as it determines the equivalent central wavelength, $\lambda_c$, at which the phase modulation by the SLM occurs. In order to describe the interference resulting from the superposition of the three spectral channels, we start with the equivalent spectrum for cSLIM imaging, namely

$$S_c(\omega) = S_i(\omega)[S_R(\omega) + S_G(\omega) + S_B(\omega)] \qquad (2)$$

where, $\omega$ is the optical angular frequency, $S_i$ is the incident light spectrum, and $S_R$, $S_G$, and $S_B$ are the spectral responses of the red, green, and blue channels, respectively. These responses capture the effect of the RGB camera color filters as well as the weighting factors [$r, g$ and $b$]. The wavelength spectra, $S_c(\lambda)$ and $S_i(\lambda)$, are shown in Fig. 1 (e), while the spectral responses of the three color channels have been plotted in Fig. S4 (c) (see Supplementary Information). The temporal autocorrelation function of the equivalent source, $\Gamma_c(\tau)$, is obtained by taking the Fourier transform of Eq. (2),

$$\Gamma_c(\tau) = \Gamma_i(\tau) \circledast [\Gamma_R(\tau) + \Gamma_G(\tau) + \Gamma_B(\tau)]. \qquad (3)$$

In Eq. (3) $\Gamma_i(\tau)$ is the autocorrelation of the illumination source and $\Gamma_R(\tau), \Gamma_G(\tau)$, and $\Gamma_B(\tau)$ are the autocorrelations corresponding to the spectra of the three color channels. The $\circledv$ symbol denotes the convolution operation and $\tau$ refers to the time delay, conjugate variable to $\omega$. A comparison between the autocorrelation functions (real part) calculated for the illumination source and the equivalent cSLIM source is shown in Fig. 1 (f). The functions are plotted against distance $d = c\tau$, with $c$ the speed of light in air.

It is apparent that the optical problem of interferometry using a color camera is equivalent to using three independent sources. In essence, although the complex fields from the three spectral channels do not interfere with one another, their autocorrelation functions add. As a result, the new, equivalent correlation function ($\Gamma_c$) is characterized by a new envelope (indicative of the coherence time, inversely proportional to the spectral width) and phase (describing the new central wavelength). The central wavelength detected in cSLIM was measured as $\lambda_c = 558\,nm$ while the incident white light was at $\lambda_i = 589\,nm$. The coherence lengths, $l_c$ and $l_i$, measured from the two corresponding autocorrelation functions [Fig. 1 (f)] were $3.24\,\mu m$ and $2.26\,\mu m$, respectively (in air). We define the coherence length as the full-width half maximum (FWHM) of the envelope [dashed line in Fig. 1 (f)] of the autocorrelation function.

The cSLIM system requires two calibration steps. First, the SLM was calibrated to ensure that the correct value of $\varphi$ was used for each of the four frames, resulting in increments of $\pi/2$ at the central wavelength $\lambda_c$ [44]. This was performed by configuring the SLM in amplitude mode and measuring the amplitude modulation in $I(x,y;\varphi)$ as a function of the SLM 8-bit grayscale input voltage [44]. The calibration curve for phase was then obtained by taking the Hilbert transform of the amplitude modulation curve [see Section S.1 of Supplementary Information]. Second, the SLIM phase reconstruction algorithm[44] includes the attenuation term $\alpha_{pc}$, measured experimentally, which is the attenuation factor of the incident light compared to the scattered light in a phase contrast objective. Since bright-field objectives do not impart this attenuation, an equivalent attenuation $\alpha_{bf}$ was introduced numerically in place of $\alpha_{pc}$ in the cSLIM phase reconstruction. The correct value of $\alpha_{bf}$ was determined by imaging an unstained tissue microarray (TMA) core using both phase contrast and bright-field objectives and tuning $\alpha_{bf}$ until the correlation between the phase values from the two acquisitions was maximized [see Section S.2 in Supplementary Information]. The resulting value of $\alpha_{bf} = 3.4$ was used for all subsequent phase reconstructions.

The typical raw outputs generated by the cSLIM system are illustrated in Fig. 2. Results are shown for an H&E stained tissue microarray at the slide [Figs. 2 (a) and (d)], core [Figs. 2 (b) and (e)] and cellular scales [Figs. 2 (c) and (d)]. The raw phase, $\phi(x,y)$, and bright-field microscopy images of the whole slide are obtained in a single scan, i.e., the bright-field image is merely one frame of the cSLIM reconstruction. These capabilities of the cSLIM scanner are also illustrated in the supplementary video file. Thus, a key feature of the cSLIM outputs is that standard histopathology and quantitative phase images are perfectly registered. This is significant in carrying out QPI studies with large cohorts, since independent pathologist evaluation can be done on the same tissue rather than a parallel section. This ensures that a pathologist's diagnosis matches precisely with markers extracted in QPI without the need for duplicate scanning of tissue section before and after staining. Thus, information from both channels (phase and bright-field) can be combined for improving results of classification and segmentation problems for which algorithms related to both modalities have been published separately so far [8,12,32,45]. The scan time for a single core (approx. 1 mm$^2$ area) was approx. 13 seconds, at a pixel ratio of 7.4 pixels/$\mu m$, which is compatible with existing commercial WSI instruments that only yield bright-field imaging [1,31].

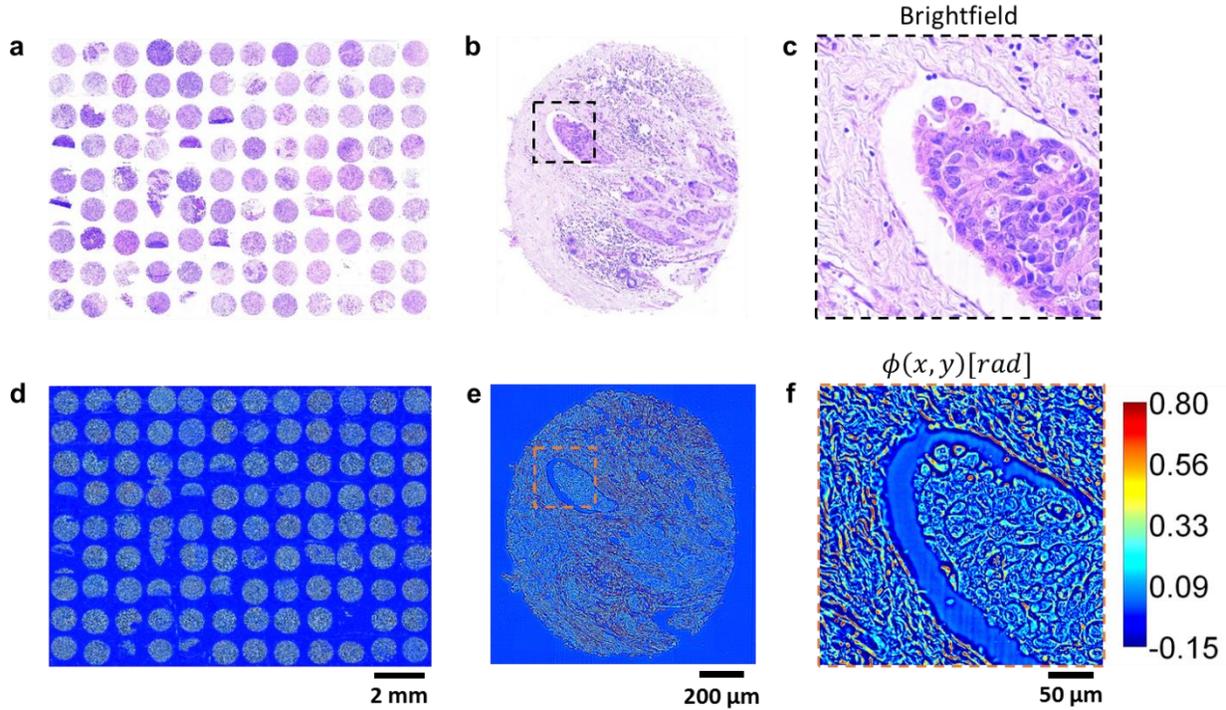

**Figure 2**. cSLIM outputs obtained by scanning a whole tissue microarray slide. **(a)-(c)** H&E stained tissue bright-field images of whole slide, single core, and epithelial region within the core, respectively. **(d)-(f)** Raw phase maps of the whole slide, single core and epithelial region within the core, respectively.

*b) Normalizing out the effects of staining*

Absorption in stained tissue is expected cause further variation in the cSLIM equivalent spectrum [shown in Fig. 1 (e)]. In addition, refractive index of stained tissue is expected to vary from that of unstained tissue because absorption and refractive index are related via the Kramers-Kronig relations [46] (see Supplementary Information, Section S.4 for analysis of dispersion in tissue). As a result, phase maps extracted from stained and unstained tissue samples may differ.

To quantify this difference, we imaged a TMA (TMA-1) of breast tissue biopsies before H&E staining using SLIM and after staining using cSLIM [see Methods, Section (a) for details about TMA-1]. The raw phase maps $\phi(x, y)$ for one TMA core, before and after staining, are illustrated in Figs. 3 (a) and (b), respectively. It is evident that staining causes a reduction in phase values as well as the image contrast. These effects are also illustrated in the histograms of the two phase images in Fig. 3 (c) where the stained tissue histogram is noticeably narrower. To normalize these effects of staining we computed for each core the standard normal variable $Z(x, y)$ from $\phi(x, y)$ using the equation

$$Z(x, y; \lambda_{c,i}) = \frac{\phi(x, y; \lambda_{c,i}) - \mu}{\sigma}, \tag{4}$$

where $\phi(x, y; \lambda_{c,i}) = 2\pi \left[ n(x, y; \lambda_{c,i}) - n_0 \right] t / \lambda_{c,i}$ is the measured phase map, with $\lambda_{c,i}$ the central wavelength of the detected spectrum $n(x, y; \lambda_{c,i})$ the tissue refractive index map, $n_0$ the refractive index of the medium surrounding the tissue and $t$ the thickness of the tissue histology section. $\mu = \left\langle \phi(x, y; \lambda_{c,i}) \right\rangle_{(x,y)}$ is the phase spatial average and $\sigma^2 = \left\langle \phi^2(x, y; \lambda_{c,i}) \right\rangle_{(x,y)} - \mu^2$ the phase spatial variance, and the operator $\langle . \rangle$ refers to the expected value, computed over spatial coordinates $(x, y)$. Both $\mu$ and $\sigma$ were computed in the tissue region only, after removal of background pixels. Details about this computation, including removal of background pixels, are described in Section S.3 in Supplementary Information.

The results of this normalization procedure are shown in Figs. 3 (d) – (f). The normalized phase maps are visually very similar between the stained and unstained tissue and the histograms seem to overlap almost perfectly. To quantify this similarity, the Pearson's correlation coefficient $\rho$ [47] was computed between the stained and unstained tissue histograms, before and after phase normalization. As illustrated in Figs. 3 (c) and (f), the $\rho$ values improved significantly due to the normalization procedure, indicating a very high degree of correlation. The $\rho$ values for a total of 30 cores (15 cancerous and 15 normal, selected randomly from TMA-1), before and after normalization, are summarized in Fig. 3 (g). The bar heights represent mean values whereas the error bars represent the standard deviations over the 30 cores. The high correlations demonstrate that $Z(x, y)$ maps are almost stain-independent, which offers excellent opportunities for quantitative pathology. Despite the stain variability associated with varying proportions of epithelium and stroma both within a core and across a population of cores, our normalization technique remains robust.

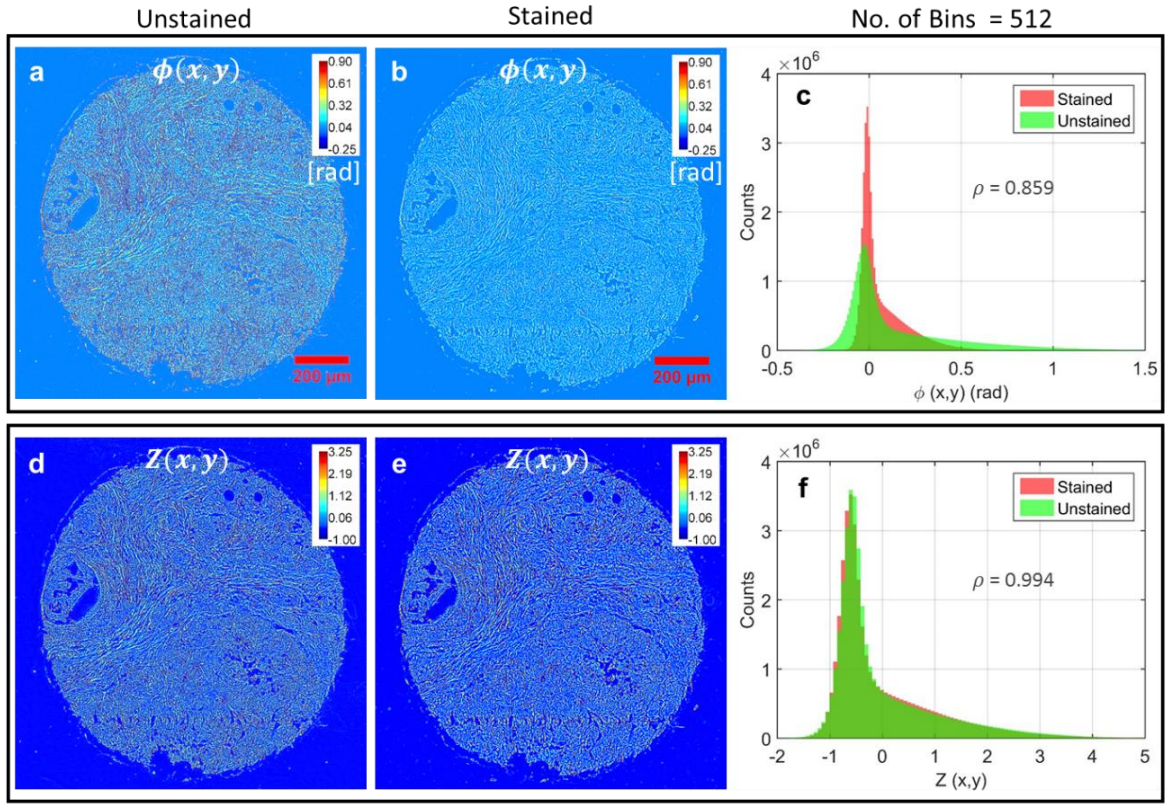
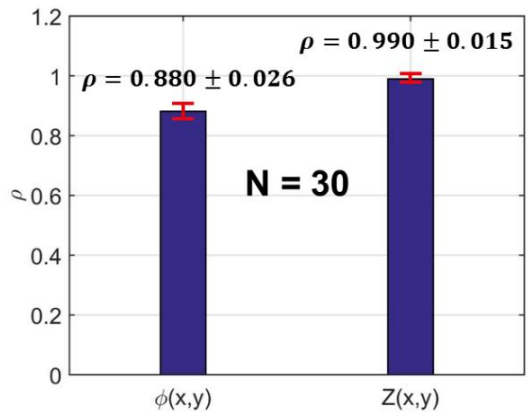

**Figure 3**. Comparison of phase maps obtained from the TMA-1 before and after staining. **(a)** and **(b)** TMA core raw phase images before and after staining, respectively. **(c)** Histograms of the phase images in (a) and (b). The similarity of the two histograms is quantified by computing the Pearson's correlation coefficient $\rho$ between them. **(d)** and **(e)** Normalized maps for the same core before and after staining, respectively. **(f)** Histograms of the images in (d) and (e). The similarity of the two histograms is quantified by computing the Pearson's correlation coefficient $\rho$ between them. **(g)** Bar plots showing the mean value of Pearson's correlation coefficient $\rho$ between core histograms for raw and normalized phase maps. Error bars show the standard deviation over the 30 cores.

It can be deduced from Eq. (4) that division by the standard deviation removes the $\lambda_{c,i}$ dependence of the phase map, minimizing the differences between unstained and stained tissue $Z(x,y)$ images caused by different spectra. While dispersion in tissue (spatial variation of both wavelength and refractive index) is also expected to cause differences between these images, our detailed analysis of this phenomenon (included in Supplementary Information Section S.4) showed that these changes are small. The fact that a global normalization is effective in removing the stain-related effects also indicates that dispersion in tissue is not a dominant effect (see Supplementary Information).

*c) Breast cancer diagnosis on stained tissue biopsies using supervised learning*

Previous reports have shown that quantitative phase images of unstained tissue can detect malignancy in different organs [29,31,32]. To demonstrate that this analysis can be extended to stained tissue cores imaged by cSLIM, we assembled $Z(x,y)$ maps both before and after staining for the 30 cores in TMA-1 selected in the previous section. Each core belonged to a different case/patient with known diagnosis results through pathologist evaluation. All malignant cases were diagnosed as infiltrating ductal carcinoma (IDC). From the $Z(x,y)$ maps, we developed a supervised learning method for breast cancer diagnosis that relied on three types of features: the median gland or epithelial region (ER) curvature $<C>$, the median of mean scattering length within an ER $<l_s>$, and the median texture vector for the ER $<T>$[32]. We used the pathologist's diagnosis for each ER as the ground truth for training [see Section (a) of Methods for details]. Details of the feature extraction, training and validation steps have been included in Methods, Section (c).

Figure 4 illustrates the diagnosis results obtained for stained and unstained tissue. Figures 4 (a), (b), (d), and (e) compare $Z(x,y)$ for benign and malignant ERs before and after staining.

Once again, the images are very similar between the two cases, proving the efficiency of our normalization procedure. In cSLIM, morphological details of these ERs are also available for traditional histopathological assessment through bright-field images [Figs. 4 (g) and (h)]. To test whether features derived from these $Z(x,y)$ maps can detect malignancy in breast tissue, 3-fold cross-validation was performed, consisting of three trials. ERs from all cores were pooled and divided into three equal sets and in each trial two sets were used for training and one set for validation. Figs. 4 (g) and 4 (h) compare the values of the three features between unstained and stained ERs, for one training set. Since texture feature $<T>$ is multidimensional [Methods, Section (c)], it is represented by its first principal component in the plot. Our results show that the feature values have a similar distribution in the unstained versus stained cases, for both normal and diseased tissue.

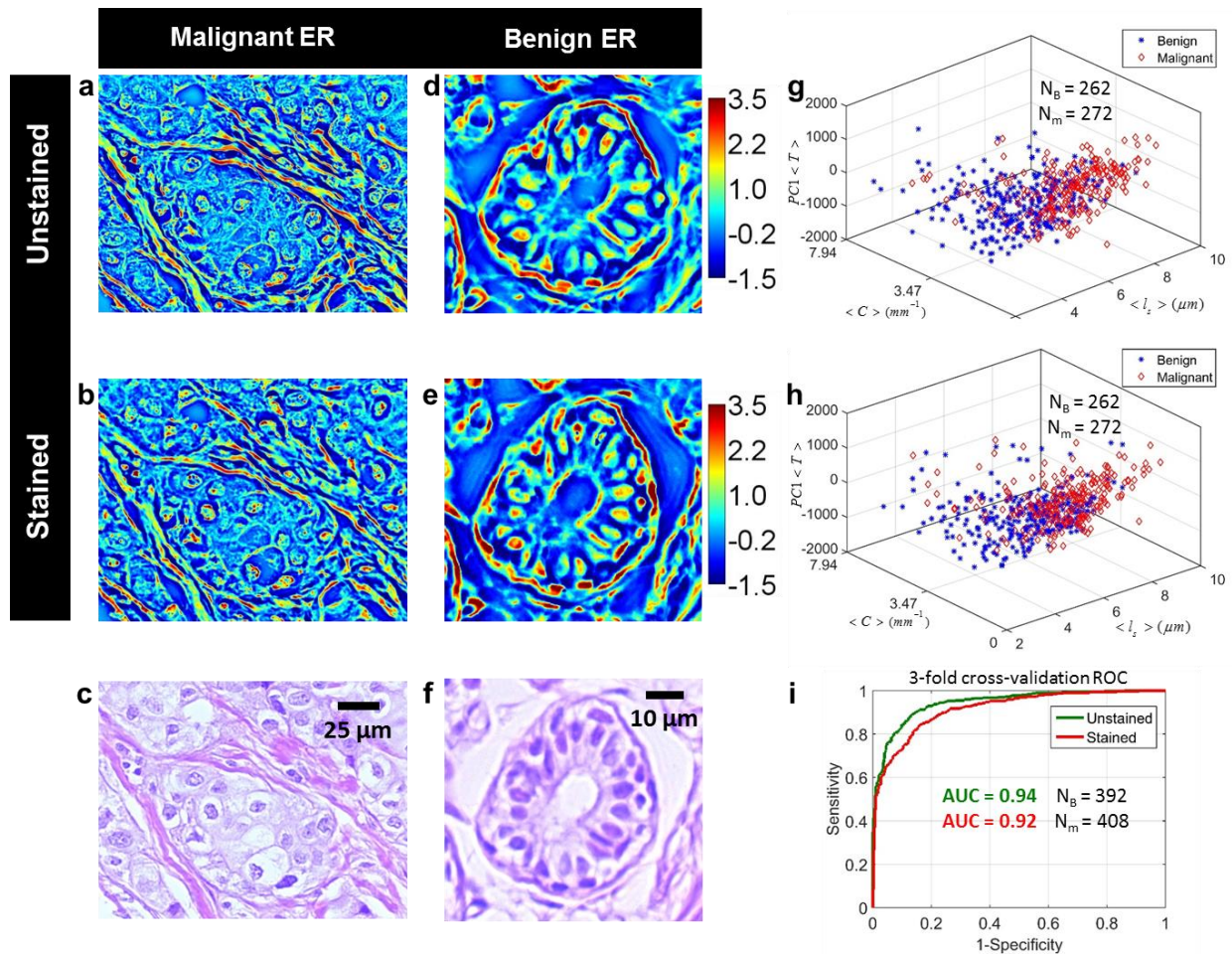

**Figure 4.** Comparison of diagnosis results between stained and unstained tissue using $Z(x,y)$ maps. **(a)-(b)** $Z(x,y)$ images of an unstained and stained malignant ER, respectively **(c)** H&E stained bright-field image of the same malignant ER. **(d)-(e)** $Z(x,y)$ images of an unstained and stained benign ER, respectively. **(f)** H&E stained bright-field image of the same benign ER. **(g)-(h)** Separation of benign and malignant ERs in training feature space for stained and unstained tissue, respectively. **(i)** ROC curves for 3-fold cross-validation for classifying benign and malignant ERs.

In each case, the probability scores for all ERs, generated by a linear discriminant analysis (LDA) classifier in all three trials, were pooled together to generate a receiver operating characteristic (ROC) curve for the cross-validation [see Methods Section (c)]. As shown in Fig. 4 (i), similar area under the curve (AUC) was measured for analysis on both stained and unstained tissue using $Z(x, y)$ maps, indicating that detection of malignancy is achievable with high accuracy using stained tissue biopsy phase maps. While the AUCs are similar, they are not identical. This can be attributed to the fact that the tissue morphology itself (while similar) is not identical between the two experiments since the process of removing the coverslip from the TMA slide and staining it results in some physical changes to the tissue biopsies.

*d) Collagen fiber orientation measurement from cSLIM data*

The prognostic value of tumor associated collagen signatures (TACS) in breast tissue has been demonstrated in a number of studies [15,48]. Traditionally, these markers have been measured using SHGM which provides chemical specificity to collagen. However, in SHGM the tumor edge is difficult to identify due to the absence of second harmonic signals in centrosymmetric structures and the acquisition speed is lower than in full-field microscopy due to the point-scanning geometry [33]. Here, we demonstrate that cSLIM can provide the relative angle $\theta$ between collagen fibers near an ER edge and the tangent to the nearest point on the edge itself [Fig. 5 (c)].

For the 30 cases selected from TMA-1 (used in the previous two sections), we measured $\theta$ in the $Z(x, y)$ maps of both stained and unstained tissue using an open source MATLAB based fiber measurement tool called CurveAlign [13] [see Methods Section (d) for details and parameter specifications]. All fibers within a distance of 63 $\mu m$ from the ER edge were considered. This is within the range of the typical intercellular signaling distance reported in literature [13,49]. ERs in all cores were segmented out before computation of $\theta$ so that their cellular structures did not interfere with the process of collagen fiber extraction [see Section (b) of Methods for details on ER segmentation].

Figure 5 compares the obtained results between unstained versus stained tissue biopsies. Figures 5 (a) and (b) show the bright-field images of malignant and benign ERs whereas Figs. 5 (c), (d), (f) and (g) illustrate the fiber orientation in their vicinity. As evident from these images, values of $\theta$ are on average larger for malignant tissue than for benign. Furthermore, the orientation measured on stained tissue qualitatively matches that measured on the unstained. Figures 5 (e) and (h) show the histograms of $\theta$ measured for all tumor adjacent fibers across the 30 core dataset. Once again malignant ERs show a greater probability of forming larger angles with their adjacent fibers. These measurements agree with previous results in literature where it was demonstrated that larger values of $\theta$ are associated with more aggressive disease and that aligned collagen fibers, oriented perpendicularly to tumor edge, facilitate local invasion [15]. The results here are significant because collagen fiber based parameters show similar values between stained and unstained normalized phase maps. Next we show that prognostic markers, based on stromal fibers, can be evaluated using cSLIM on tissues from cancer patients with different outcomes.

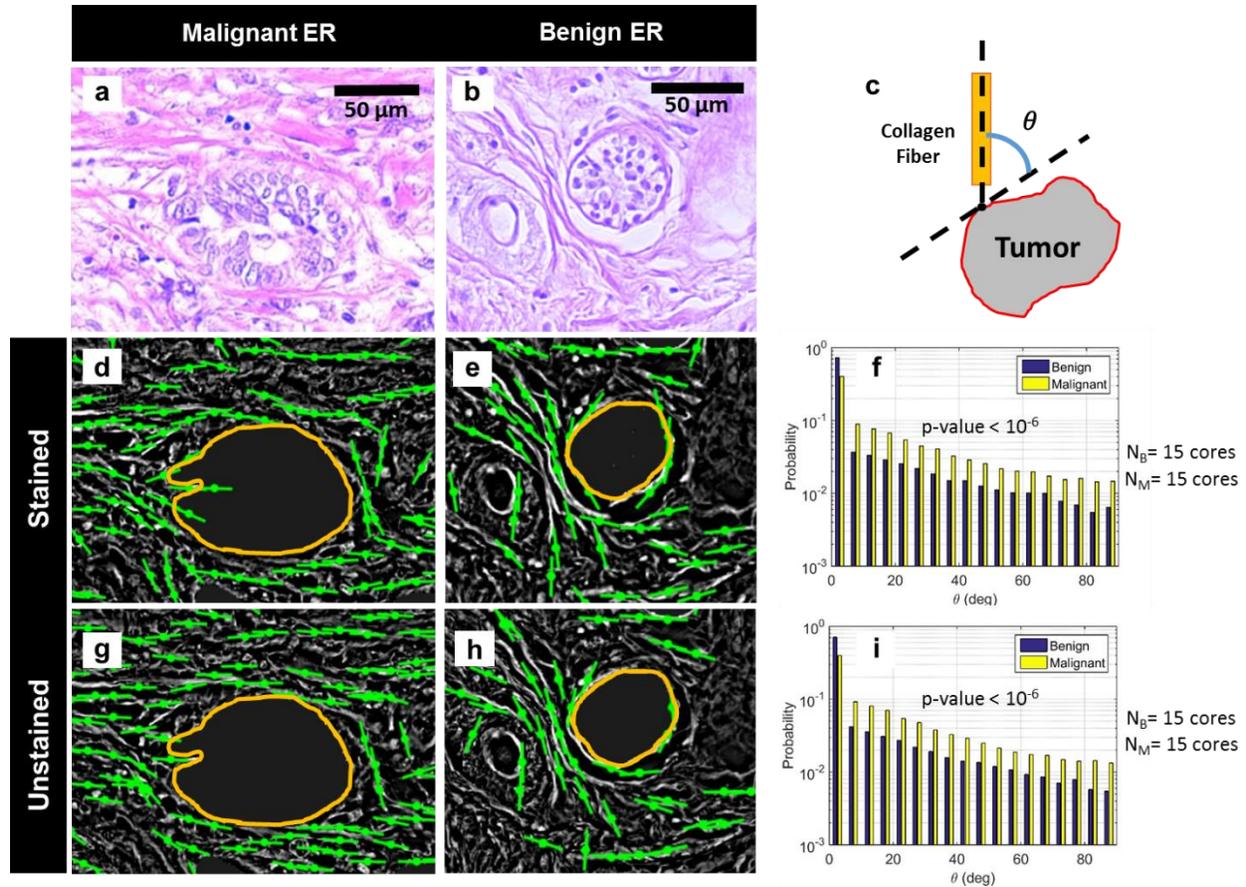

**Figure 5.** Comparison of relative collagen fiber angle $\theta$ between stained and unstained tissue biopsies. **(a)** and **(b)** Bright-field images of malignant and benign ERs, respectively. **(c)** Illustration of relative fiber angle $\theta$. **(d)** and **(e)** Orientations of collagen fibers in stained tissue, in the vicinity of malignant and benign ERs, respectively. Fiber orientations are shown as green lines while the ER edge is marked in orange. **(f)** Normalized histogram of $\theta$ measured for all ERs within the stained tissue dataset (15 malignant and 15 benign cores). The p-value for separation between the two classes was determined using the two-sample Student's T-test. **(g)** and **(h)** Orientation of collagen fibers in unstained tissue in the vicinity of malignant and benign ERs, respectively. Fiber orientations are shown as green lines while the ER edge is marked in orange. **(i)** Normalized histogram of $\theta$ measured for all ERs within the unstained tissue dataset (15 malignant and 15 benign cores). The p-value for separation between the two classes was determined using the two-sample Student's T-test.

*e) Quantifying aligned collagen fibers in stained tissue for prognosis*

Using the fiber extraction and orientation measurement demonstrated in the previous section, here we measure the prognostic marker TACS-3 [15]. TACS-3 refers to the presence of aligned collagen fibers that terminate at the tumor edge at high (near perpendicular) values of relative angle $\theta$. In Bredfeldt *et al.* [13], TACS-3 was measured using an automated, supervised learning scheme within a TMA imaged using SHGM. Using cSLIM, we imaged the same H&E stained TMA (TMA-2) and extracted $Z(x,y)$ maps for each core. We then used the fiber analysis method described earlier [13] to demonstrate that cSLIM can successfully detect TACS-3.

TMA-2 comprised 196 cases (1 core per case) of IDC with disease free survival (DFS) and disease specific survival (DSS) information available for each case [15]. Using open-source MATLAB based fiber quantification tools CT-FIRE[13,50,51] and CurveAlign, features were computed for each fiber that was within a distance of 100 $\mu m$ from the tumor edge [see Section (e) of Methods for details on fiber extraction and feature computation]. Once again, this distance was chosen to be comparable with the typical intercellular signaling distances reported in literature [49]. The features extracted from each fiber were then combined to generate a feature vector for the core. Three core-level features were found to be most informative in distinguishing between TACS-3 positive and negative patients: mean of $\varepsilon$ (mean nearest fiber alignment), mean of $l$ (nearest distance of fiber from tumor edge), and skewness of $\theta$. These features are listed in the table of Fig. 6 (a). For detailed feature description see Methods, Section (e) and Refs. [13,50].

Three-dimensional feature vectors for 10 cores marked as TACS positive and 10 cores marked as TACS-3 negative (based on pathologist consensus [15]) were used as predictors for training a linear Support Vector Machine (SVM) classifier. The classifier was then used to classify all 196 cores as either TACS-3 positive or TACS-3 negative. Figure 6 (a) shows the difference in the feature means of groups classified as TACS-3 positive and TACS-3 negative. According to these mean values, cores classified as TACS-3 positive have a higher probability of containing aligned fibers (high $\varepsilon$) that terminate at or near the tumor edge (low $l$). Furthermore, these cores have $\theta$ histograms that are more positively skewed, reflecting an asymmetry due to more instances of high $\theta$. These measurements agree with the pathologist definition of TACS-3, indicating successful classification.

Survival analysis was carried out to test whether patients classified as TACS-3 positive had significantly worse outcomes than those deemed TACS-3 negative. Figure 6 summarizes these results. Univariate Cox proportional hazard regression [52] and Kaplan-Meier estimates [52] were used to compare survival between TACS-3 positive and TACS-3 negative cases. As shown in Fig. 6 (b), TACS-3 positive patients had hazard ratios[52] of greater than 2 and p-values < 0.05, representing a statistically significant chance of worse disease specific survival (DSS) and disease free survival (DFS) outcomes (see Methods Section (a) for definitions of DSS and DFS). This trend is also evident in the Kaplan-Meier estimate of the DFS and DSS survival functions [Fig. 6 (d)] where TACS-3 positive patients show significantly higher frequency of events. The p-values in this case were computed using the log-rank test [53]. Finally, we also computed the Pearson's correlation coefficient $\rho$ between the computationally generated TACS-3 scores using the cSLIM images and manual scores generated by pathologists (see Ref. [15] for details) . As shown in Fig. 6 (d), a positive correlation was measured in each case, indicating that the automated analysis extracted similar histological markers as were observed by the pathologists manually.

Our results are significant because in addition to quantitating the TACS-3 marker, the cSLIM system provides high-throughput acquisition, operates on existing stained tissues, and provides simultaneous bright-field/color images. The instrument, thus, allows the assessment of traditional prognostic markers (e.g. tumor grade and molecular subtype) as well as new prognostic markers (such as TACS-3) in a single scan. This advantage is obtained at the expense of some additional instrument optics with no additional constraints on sample preparation.

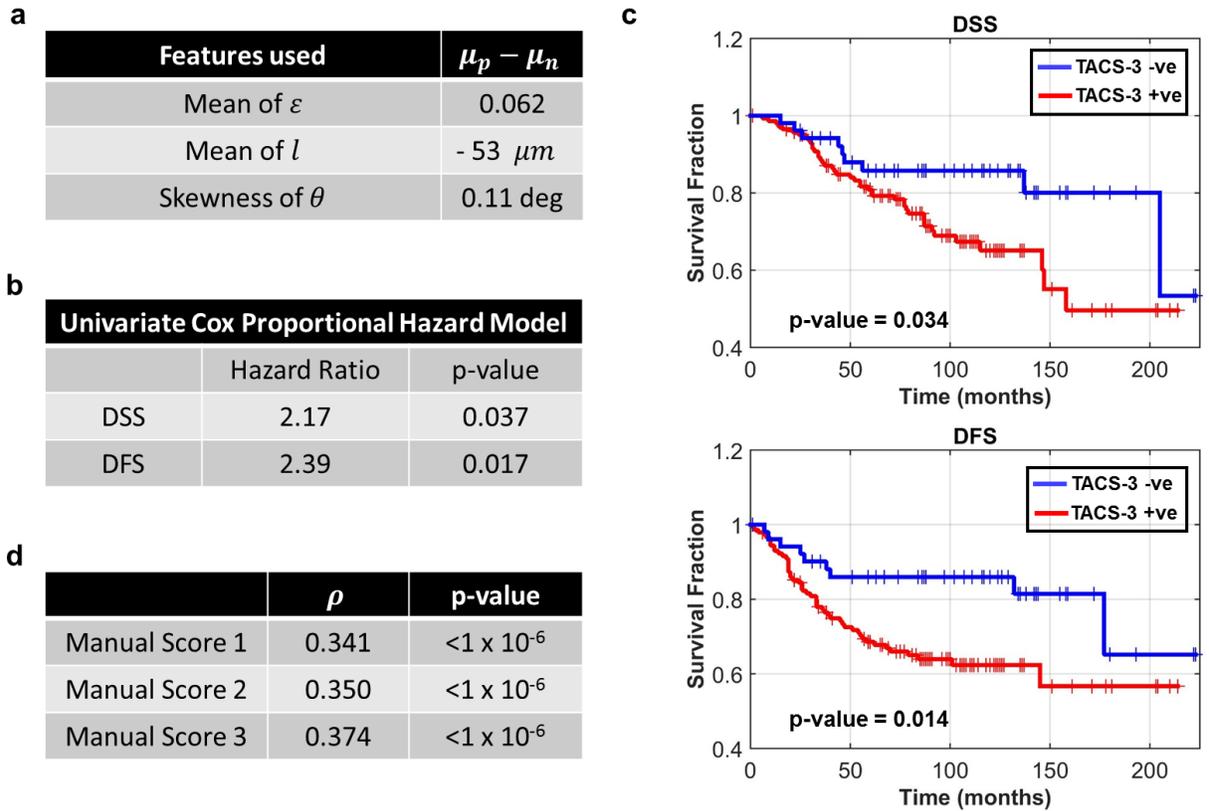

**Figure 6.** Results of survival analysis for TACS-3 positive and TACS-3 negative groups, classified based on fiber features measured from $Z(x, y)$ maps obtained by cSLIM. **(a)** Fiber features found to be most informative in classifying cores as positive or negative. $\mu_p$ and $\mu_n$ refer to the mean of these features for the groups classified as positive and negative, respectively. **(b)** Results of the Univariate Cox proportional hazard regression with TACS-3 status as variable. **(c)** Kaplan-Meier estimate of survival function for DSS and DFS. Vertical tic marks represent right-censoring events. The number of cores classified as TACS-3 negative were 52 and those classified as TACS-3 positive were 144. **(d)** Pearson's correlation between the automated TACS-3 scores and the 3 different scores generated manually by pathologist evaluation, as outlined in Ref.[15].

## 3. Summary and Conclusion

In summary, we presented cSLIM, a tissue imaging modality that provides stain-independent, quantitative markers of disease while simultaneously providing traditional histopathology images. Our instrument facilitates clinical translation of our disease markers by posing no new sample preparation requirements and providing traditional and novel markers in a single acquisition. These advantages are obtained by simply upgrading an existing microscope to a phase-sensitive instrument with an optics module at the output port. Because cSLIM provides full-field diffraction-limited imaging with visible light, it results in much higher resolution and throughput with relatively inexpensive equipment, compared to other label-free methods (vibrational spectroscopy and non-linear microscopy). We demonstrated that cSLIM normalized phase maps are stain-independent by comparing the results of phase imaging of tissue before and after staining. We also demonstrated that diagnostic and prognostic markers relevant to breast cancer can be extracted from cSLIM images through automated machine learning techniques. Since these disease markers are quantitative, they eliminate subjectivity from tissue evaluation. Because the analysis is automated, disease markers can be obtained by pathologists rapidly and reproducibly. Furthermore, by providing a fast and convenient means of extracting stromal collagen based prognostic markers, our instrument can potentially help pathologists predict disease aggressiveness in patients for whom other more traditional disease markers fail.

While we focused here on breast diagnosis and prognosis, it is evident that cSLIM can contribute to the pathology of other diseases. Furthermore, imaging tissues stained with other agents besides H&E is also feasible with cSLIM without modifications. For these reasons, we anticipate that cSLIM will have a low threshold for adoption at large scales.

## 4. Methods

*a) Tissue microarrays*

TMA-1 was purchased unstained from US Biomax (Serial # BR1002) and comprised cases of IDC and normal benign tissue. The TMA was obtained from the manufacturer with all human subject identifiers removed. Neither the authors nor their institutions were involved in the tissue collection. Details regarding this TMA have already been reported in Ref. [32]. Case-by-case diagnosis was provided by the manufacturer's board certified pathologist through examination of both H&E stained tissue and immunohistochemical (IHC) makers, both on parallel sections of tissue. After acquiring unstained tissue phase images, the TMA was H&E stained for cSLIM imaging using standard protocols [54]. Before staining, the coverslip was removed from the slide and post-staining the slide was re-coverslipped using the same mounting medium as before (Xylene). As discussed in Results and Discussion, 15 cases of IDC and 15 cases of normal tissue were randomly selected from the TMA for the studies done in this paper. One core per case was available. Each biopsy core had a diameter of 1 mm and a thickness of 5 $\mu m$. For the IDC cases, a second board certified pathologist also marked any benign regions within the tissue cores which were then excluded from the analysis. In this way, diagnosis of each ER within each core (Benign or Malignant) was available.

TMA-2 was used in previous studies by Bredfeldt *et al.* [13] and Conklin *et al.*[15]. Details regarding patient profiles, tissue processing and core selection have been described in Ref. [15]. The dataset used from the TMA consisted of 196 cores (1 core per patient) and patients were followed up for a median time of 6.2 years, ranging from 1-223 months in order to determine patient outcomes. DSS and DFS information was available for each patient. DSS was defined as the time from diagnosis to death from breast cancer or date of last follow up evaluation. DFS was defined as the time from date of diagnosis to the first date of recurrence. TMA-2 was H&E stained using standard protocols [54], allowing for simultaneous acquisition of both normalized phase and bright-field images using the cSLIM system. All tissue and patient information were obtained after approval by Institutional Review Board (IRB) [15].

*b) Epithelial tissue segmentation for feature extraction*

For computation of epithelial features during supervised learning as well as for measurement of relative fiber orientation, knowledge of the ER boundary is required. For TMA-1, the ER boundaries were annotated in all the cores manually using the region-of-interest tool in ImageJ by using the H&E stained tissue bright-field images as a guide. Consistent criteria were used during annotation– groups of epithelial cells bounded by stroma on all side were considered a single ER. Other tissue components were considered part of the ER if surrounded on all sides by epithelial cells [32].

For TMA 2 we used the same ER segmentation masks as in Ref. [13]. We registered the bright-field images from cSLIM with those from the original study by using Speed-Up Robust Features (SURF) [55]. The $Z(x, y)$ maps were, thus, registered with the ER segmentation masks as well.

*c) Classification scheme for cancer diagnosis*

The supervised classification of benign versus malignant lesions [Section (c) of Results and Discussion] is based on the procedure we reported in Ref. [32]. We apply this procedure to our $Z(x, y)$ images, in three steps: feature extraction, training, and validation.

During feature extraction, first maps of the ER curvature $C$, mean scattering length $l_s$ and texture vector $T$ were extracted for each core within our datasets. The ER curvature $C$ refers to the extrinsic curvature of a two-dimensional plane (in this case a benign or malignant ER) and can be construed as the magnitude of the rate of change of a vector tangent to the ER perimeter. We used an open-source MATLAB code to measure $C$ for each annotated ER [56]. The mean scattering length $l_s$ is the length-scale over which a single scattering event happens on average and can be computed from tissue phase images using the scattering-phase theorem [57]. The texture vector $T$ consists of frequencies of elements known as 'textons' within the vicinity of a pixel in the image. Textons have been shown to be effective measures of the unique texture surrounding a pixel [12,58,59]. To separate benign and malignant lesion in this paper, contrary to our previous work where 50 textons were trained (resulting in a 50 dimensional vector $T$) [32], we found that 30 textons were sufficient due to the smaller size of the dataset. This number was obtained iteratively by measuring the cross-validation AUC (discussed below) while increasing the number of textons and stopping at the point where no improvement in AUC was noticed, to avoid overfitting. For feature extraction, all other parameters were identical to those used in Ref. [32].

After pixel-wise computation of these features, the median over each ER was calculated for each feature $[<C>,<l_s>,<T>]$, using the ER masks obtained through manual segmentation [Methods, Section (b)]. These features were then concatenated to generate an overall (32 dimensional) feature vector for each ER. Using pathologist diagnosis result for each ER as the class label (benign or malignant) and its overall feature vector as the predictor, an LDA classifier was trained. During validation, feature vectors for unknown ERs were input to the classifier which generated likelihood scores for the output classes. During validation, the overall data set was partitioned into three equal partitions. Three validation trials were performed (3-fold cross validation) [60]. In each trial, two partitions were used for training and the remaining one is for validation. The classifier performance was measured using ROC curve analysis. Likelihood scores for each ER, generated by the classifier from the three validation trials, were pooled together [61] to generate an overall ROC curve [Figs. 4 (i)] and the AUC was used as a metric for classifier accuracy.

*d) Fiber orientation extraction on TMA-1 using CurveAlign*

In Results and Discussion Section (d), we compared $\theta$, the relative angle between the orientation of a collagen fiber and the tangent to the nearest point on the tumor edge [depicted in Fig. 5 (c)], between benign and malignant cases. The results were extracted using an open source MATLAB based tool called CurveAlign, algorithmic details of which have already been described in a number of publications [13,50,62,63]. For our analysis in Section (d), we chose the Curvelet Transform (CT) based fiber analysis method within CurveAlign. This method uses the curvelets provided by curvelet transformation [64] of the image to represent the edges of collagen fibers, without segmenting the individual fibers. From computation of these curvelets, thus, scale, location and relative orientation of each fiber can be calculated [50]. TIFF files that contained masks of the ERs [obtained through manual annotation, described in Methods Section (b)] were used in the 'Boundary Method' field within CurveAlign. The fraction of coefficients to keep, during curvelet transform computation, was set at 0.005 and the distance from the tumor edge, up to which fibers are analyzed, was set to 100 pixels or approx. 63 $\mu m$. Before extraction of $\theta$, ERs were segmented out from all the core images so that the cellular structures within them did not interfere with the process of fiber extraction during curvelet transformation.

*e) TACS-3 measurement on TMA-2 using CT-FIRE and CurveAlign*

As described in Results and Discussion, for detecting the TACS-3 prognostic marker on cSLIM $Z(x,y)$ maps the same general method as used in Ref. [13] was employed. We summarize that analysis method here. First, the ERs within each core were segmented out using its corresponding segmentation mask [see Section (b) in Methods]. This step was carried out to ensure that during subsequent fiber segmentation, cell edges did not interfere. CT-FIRE was then used to segment out all fibers within the $Z(x,y)$ map of each core. Default parameters, as outlined in the CT-FIRE manual [51], were used except for the parameters labelled "thresh_im2" and "s_xlinkbox" for which values were set to 30 and 5 respectively. These fiber segmentation maps, along with ER segmentation masks, were then input to CurveAlign for extraction of fiber features. Features were computed for all fibers that were within a distance of 100 $\mu m$ from the tumor edge. In CurveAlign, the "CT-FIRE Fibers" fiber analysis method was chosen and the "TIFF Boundary" was chosen as the boundary method. CurveAlign extracts a total of 34 fiber features as part of its standard computation. These features are related to the fiber curvature, width, length, density, alignment, proximity to epithelium and relative angle to epithelial boundary [13]. As discussed in the main text, core-level statistics derived from 3 features (related to alignment, proximity to epithelium and relative angle) were found to be the best predictors of DSS and DFS. The first of these features was the mean nearest alignment $\varepsilon$, defined as the mean of the alignment of a fiber to its nearest 2, 4, 8 and 16 fibers. The algorithm for measuring the alignment of a fiber to its nearest neighbors has already been described in Ref. [13]. The second feature used was the distance of each fiber to the nearest ER boundary, denoted $l$. The final feature was the relative fiber angle $\theta$. Means of $\varepsilon$ and $l$ and skewness of $\theta$, over all fibers in each core, were used as predictors for SVM training [Section (e) of Results and Discussion]. For survival analysis, the Kaplan-Meier estimate was computed using an open-source MATLAB code [65].

**Data Availability:**

The datasets generated or analyzed during this work are available from the corresponding author upon reasonable request.

**Acknowledgements:**

This work was supported by the National Science Foundation (CBET-1040461 MRI, CBET-0939511 STC, DBI 1450962 EAGER and IIP-1353368). This work is also supported by the National Institutes of Health (K. E) under grant# R01CA206561. H. M's PhD thesis research is being supported by the Beckman Graduate Fellowship Program administered by the Arnold and Mabel Beckman Foundation.

**Competing interests:**

G. P has financial interest in Phi Optics, Inc. a company that develops quantitative phase imaging technologies. K. E is co-founder of OnLume Inc., a company that develops imaging solutions for fluorescence image guided surgery.

**Author contributions statement:**

H. M and G. P conceived the design of the imaging system. H. M imaged the tissue samples and performed the image processing and analysis. A. K and Y. L were involved in the image processing and analysis for the study on TACS-3 detection. K. E supervised the study on TACS-3 detection. M. K developed the tissue scanner used for imaging the samples. T. N contributed to the image processing and analysis for cancer diagnosis. A. B and K. T provided clinical pathology expertise. H.M. and G. P wrote the manuscript and all other authors contributed edits. G.P. supervised the project.

**Materials and Correspondence**

Address material requests and other correspondence to G.P, Email: gpopescu@illinois.edu.

**Quantitative Histopathology of Stained Tissues using Color Spatial Light Interference Microscopy (cSLIM): Supplementary Information**

**S.1 SLM calibration with RGB camera**

SLIM requires SLM calibration so that the correct phase modulation is applied between the scattered and incident components of light [1]. As shown in Figs. 1 (e) and (f) in the main text, the spectrum of incident light is different from the equivalent cSLIM spectrum in both spectral width and central wavelength. For these reasons, in contrast with grayscale imaging, recalibration of the SLM was required for the cSLIM system. After configuring the SLM in amplitude modulation mode [1], the 8-bit grayscale input to the SLM was scanned from 0-255. The corresponding amplitude modulation in $I(x, y; \varphi)$ [given by Eq. (1) in the main text] was measured and each frame was averaged to generate a one dimensional amplitude modulation curve, shown in Fig. S1 (a). By taking the Hilbert transform of this curve, the SLM calibration curve [Fig. S1 (b)], relating phase values to grayscale input, was obtained. The grayscale values corresponding to $\varphi = 0, \pi/2, \pi, 3\pi/2$ rad were used in all imaging experiments.

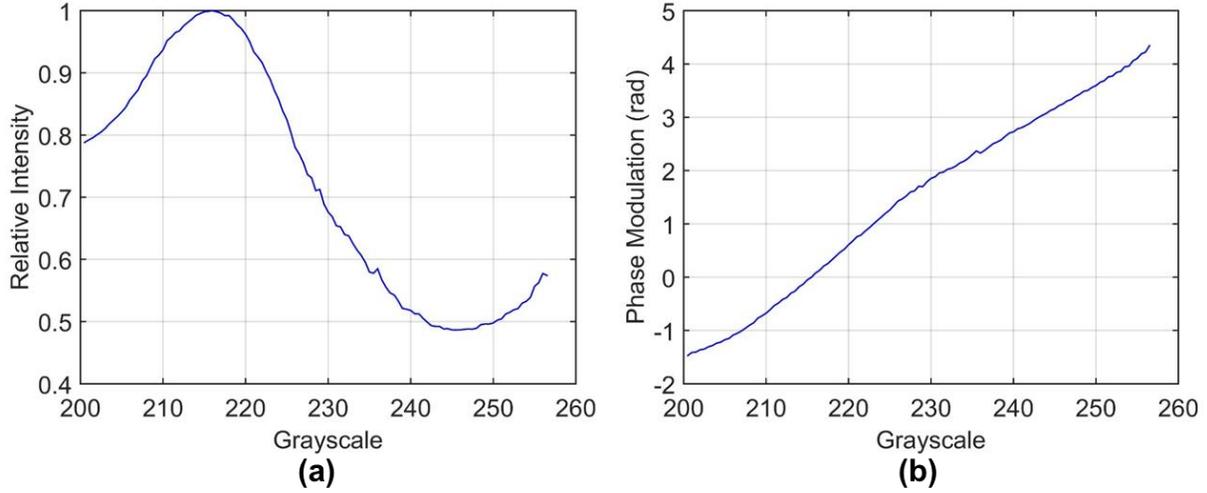

**Figure S1.** (a) Amplitude (intensity) modulation curve obtained in cSLIM by configuring the SLM in amplitude mode. (b) Phase calibration extracted using the Hilbert transform of the curve from (a).

### S.2 Attenuation calibration for the bright-field objective

Phase contrast objectives contain a phase plate in their back focal planes that not only introduces a $\pi/2$ rad phase shift between the unscattered light $U_0$ and scattered light $U_1$ but also imparts an attenuation factor $\alpha_{pc} = \frac{|U_1|}{|U_0|}$. This factor is used during SLIM phase reconstruction while calculating the ratio of the amplitudes of the two interfering fields [1,2]. In our cSLIM experiments, we have used a bright-field objective that does not have this attenuating element. An equivalent value of attenuation, $\alpha_{bf}$, was, therefore, used during phase reconstruction. $\alpha_{bf}$ was obtained through the following calibration procedure, which is illustrated in Fig. S2. An unstained TMA core was imaged using cSLIM with both a phase contrast and a bright-field objective. During phase reconstruction for the phase contrast case, the measured attenuation factor $\alpha_{pc} = 1.97$ was used and the raw phase image $\phi(x, y)$ was obtained [Fig. S2 (c)]. For the bright-field case, the equivalent attenuation $\alpha_{bf}$ was numerically tuned and the phase $\phi(x,y)$ was obtained for each $\alpha_{bf}$.
For each $\alpha_{bf}$, we calculated the cross-correlation $\gamma(P_{pc}, P_{bf})$ between the probability distributions $P_{pc}$ and $P_{bf}$ of $\phi(x,y)$ in the phase contrast and bright-field cases, respectively [3]. $P_{pc}$ and $P_{bf}$ were obtained by normalizing their respective image histograms (constructed with 512 bins each). As shown in Fig. S2 (b), $\gamma(P_{pc}, P_{bf})$ maximizes at $\alpha_{bf} = 3.4$, which was, thus, the value used for all subsequent imaging experiments. Fig. S2 (d) shows the phase image obtained using the bright-field objective at $\alpha_{bf} = 3.4$. It has similar phase values compared to those obtained using a phase contrast objective [Fig. S2 (c)].

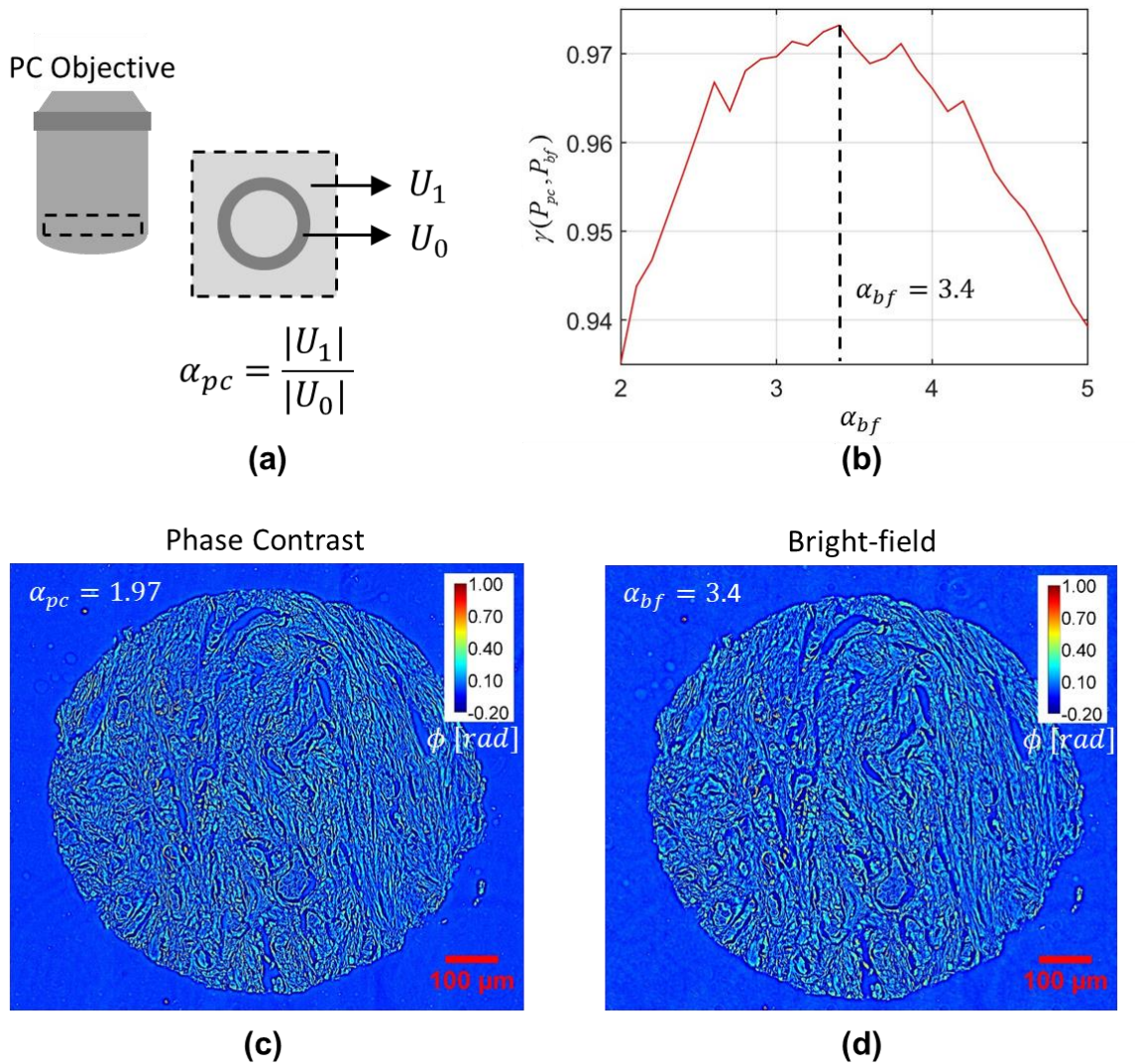

**Figure S2.** (a) Attenuation $\alpha_{pc}$ of the unscattered light $U_0$ with respect to scattered light $U_1$ in a phase contrast objective. For imaging with a bright-field objective this term is introduced numerically, $\alpha_{bf}$. (b) Cross-correlation coefficient $\gamma$ between the probability distributions $P_{pc}$ and $P_{bf}$ of the same unstained TMA core measured using phase contrast and bright-field objectives, respectively. $\gamma$ is measured as a function of $\alpha_{bf}$ and peaks at $\alpha_{bf} = 3.4$. (c) cSLIM image of the unstained tissue core obtained using a phase contrast objective with the measured value of attenuation $\alpha_{pc} = 1.97$. (d) cSLIM image of the unstained tissue core obtained at the optimum value of $\alpha_{bf} = 3.4$.

## S.3 Procedure for stain normalization

As discussed in Results and Discussion Section (b) of the main text, we extracted the normalized phase maps $Z(x, y)$ from the raw phase maps $\phi(x, y)$ generated by the cSLIM system, using Eq. (4).

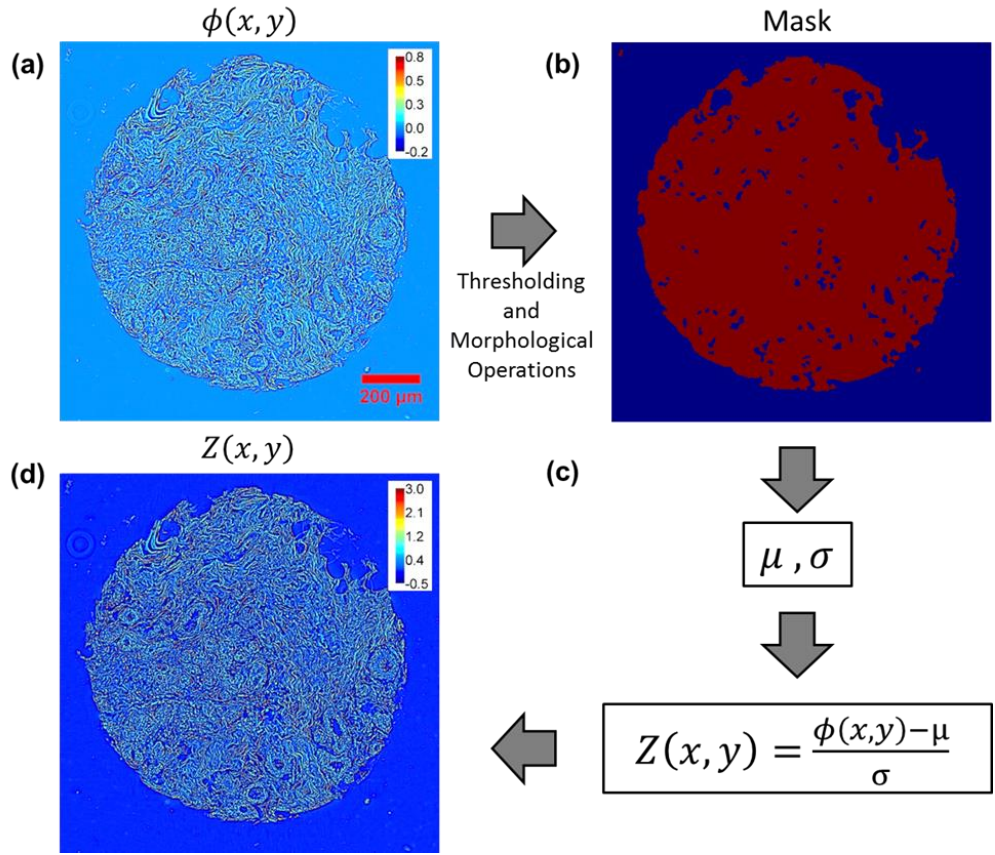

**Figure S3.** **(a)** Raw phase map of stained tissue microarray core. **(b)** Mask for computation of $\mu$ and $\sigma$ over foreground pixels. **(c)-(d)** Computation of normalized phase map.

This computation is illustrated in Fig. S3. For each tissue core $\phi(x, y)$ map [Fig. S3 (a)], we first generated a segmentation mask [Fig. S3 (b)] using thresholding followed by morphological operations (closing and removal of all 8-neighboring connected objects with total area smaller than 1600 $\mu m^2$). A circular structural element was used for morphological closing with a diameter of approx. 4 $\mu m$. The mask was then used to calculate the mean $\mu$ and standard deviation $\sigma$ of the foreground region in $\phi(x,y)$ (region occupied by tissue core). Finally, the $Z(x,y)$ image was computed for the core from these parameters as shown in Figs. S3 (c) and (d).

**S.4 Analysis of dispersion in stained tissue**

As discussed in the main text, the incident light and cSLIM spectra are different due to the effect of the spectral responses of the red, green, and blue channels of the cSLIM system. This results in different values of both the central wavelength and the coherence length for RGB and grayscale imaging even for an unstained tissue sample [see Figs. 1 (e) and (f) in the main text]. In the presence of H&E stained tissue, these two spectral parameters are expected to change even further due to absorption in dye. Our results suggest that normalization of both stained and unstained tissue phase maps [using Eq. (4) in the main text] accounts for these differences, making results from both cases very similar. We detail in this section reasons for why this normalization removes the stain dependent signal from phase images.

To explore this, we first compute the pixel-wise spectra of light detected by the RGB camera when an H&E stained tissue core is imaged. From the bright-field image measured by the camera [illustrated in Fig. S4 (a)], we extract the red, green and blue channel intensity images, $R(x,y), G(x,y)$, and $B(x,y)$, respectively. By dividing each image by the average signal in a 30 x 30 pixel background region within it (where no tissue is present), we are able to obtain the transmission images for three channels: $t_R(x,y), t_G(x,y)$ and $t_B(x,y)$ [Fig. S4 (b)]. Fig. S4 (c) shows the spectral responses, in the absence of tissue absorbance, of the red, green and blue channels [$S_R(\lambda), S_G(\lambda)$ and $S_B(\lambda)$, respectively]. These responses include both the filter responses of the camera as well as the weights attached to the three channels numerically during computation of the equivalent grayscale image $I(x,y)$ [Eq. (1) in the main text]. Thus, in the *absence* of tissue absorbance, the spectrum for cSLIM imaging, $S_c(\lambda)$, can be computed as

$$S_c(\lambda) = S_i(\lambda)[S_R(\lambda) + S_G(\lambda) + S_B(\lambda)], \tag{S1}$$

where $S_i(\lambda)$ is the incident light spectrum. $S_i(\lambda)$ and $S_c(\lambda)$ are plotted in Fig. 1(e) of the main text.

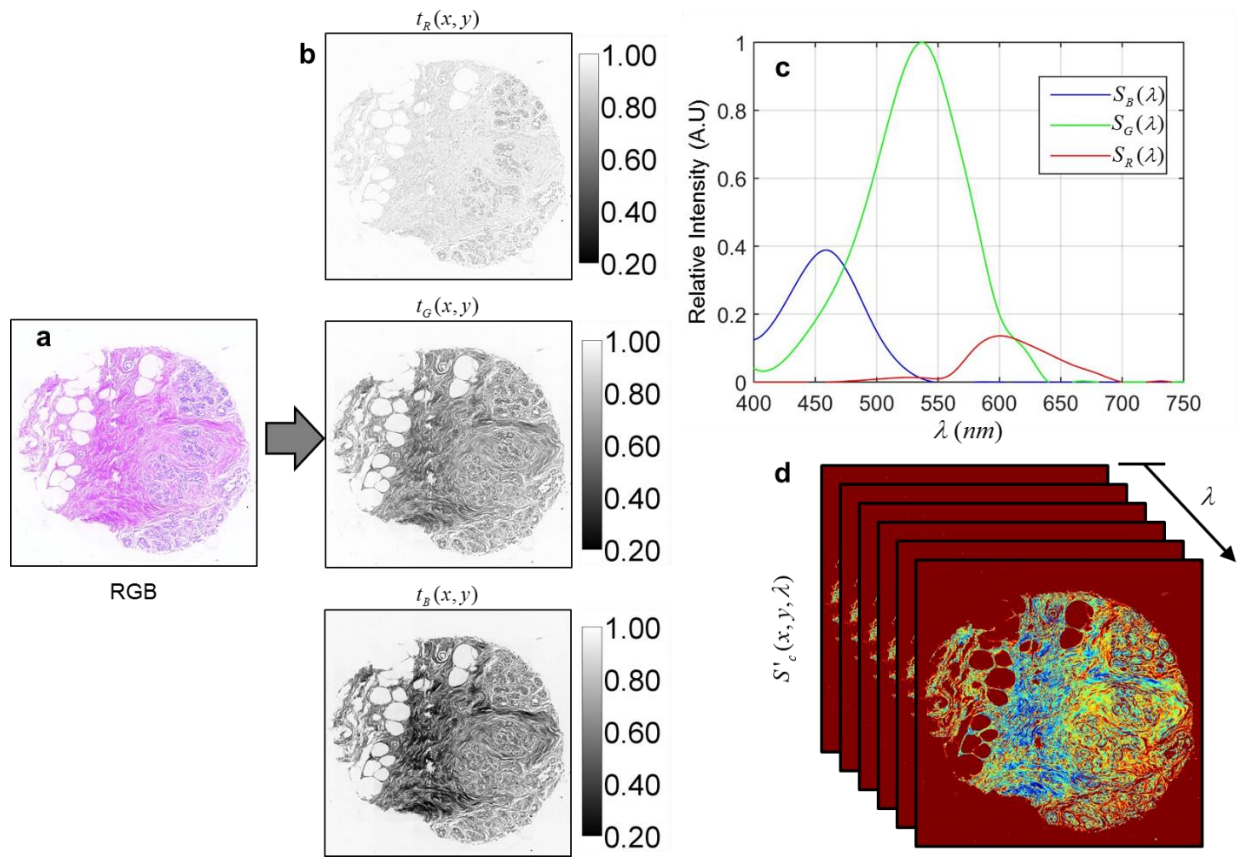

**Figure. S4 (a)** Bright-field image of a stained TMA core. **(a)** Transmission maps for the red, green and blue channels. **(c)** Spectral response of cSLIM for the red, green and blue channels in the absence of tissue absorbance. **(d)** cSLIM spectrum at each pixel computed using Eq. (S2).

Assuming constant values of transmission for each spectral channel per pixel [$t_R(x,y)$, $t_G(x,y)$, and $t_B(x,y)$] the spectral response for each channel in the *presence* of tissue absorbance can be calculated as $S'_j(x,y,\lambda) = S_j(\lambda)t_j(x,y)$ where $j = R, G, B$. Multiplying each response by the illumination spectrum $S_i(\lambda)$ and summing them gives us the pixel-wise cSLIM spectrum in the presence of tissue absorbance, namely

$$S'_c(x,y,\lambda) = S_i(\lambda)[S'_R(x,y,\lambda) + S'_G(x,y,\lambda) + S'_B(x,y,\lambda)]. \tag{S2}$$

This spatially-resolved spectrum is illustrated in Fig. S4 (d). From $S'_c(x,y,\lambda)$ we can compute the central wavelength map $\lambda'_c(x,y)$ for the stained tissue core. By subtracting the mean value of $\lambda'_c$ in a 30 x 30 pixel background region from $\lambda'_c(x,y)$, we obtain the map of the wavelength shift due to tissue absorbance, $\delta\lambda'_c(x,y)$, illustrated in Fig. S5 (a). This image shows that dispersion (variation of $\lambda'_c$ in $x$ and $y$) due to tissue absorbance is small. To quantify this effect, we calculated the histogram (computed over the foreground region consisting of tissue only) in Fig. S5 (b) which shows a mean shift of 2.3 nm with a standard deviation of 1.5 nm. Absorbance related spectral changes also cause modifications in the coherence length $l'_c$ across the stained tissue. The image $l'_c(x,y)$ can be obtained from $S'_c(x,y,\lambda)$ by computing its autocorrelation function through Fourier transformation and measuring the full-width half maximum (FWHM) of the function's envelope [see Results and Discussion, Section (a) of the main text]. The shift in coherence length $\delta l'_c(x,y)$ can be extracted from $l'_c(x,y)$ by subtracting from it the mean value of $l'_c$ in a 30 x 30 pixel background region. $\delta l'_c(x,y)$ is shown in Fig. S5 (c) and its histogram is shown in Fig. S5 (d). We measured a mean of $-81.0\,nm$ and a standard deviation of $58.3\,nm$ for $\delta l'_c(x,y)$.

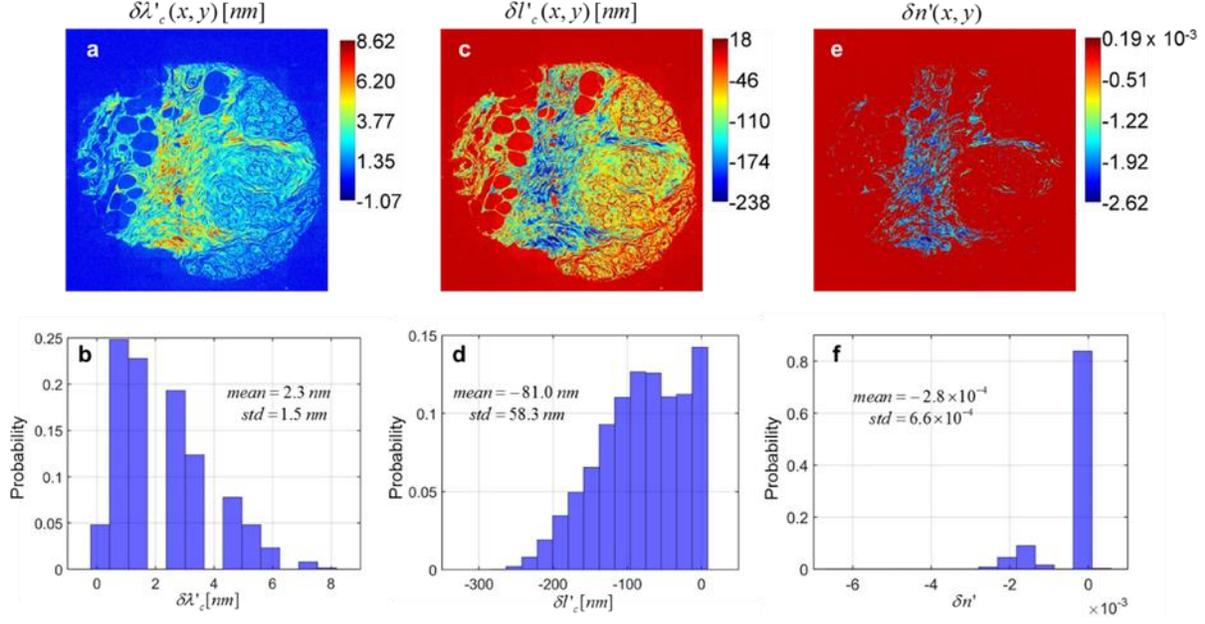

**Figure S5. (a)** Central wavelength shift across the tissue core due to dispersion. **(b)** Normalized histogram of the image in (a) over the foreground (tissue region). **(c)** Shift in coherence length across tissue core due to dispersion. **(d)** Normalized histogram of the image in (c) over the foreground (tissue region). **(e)** Dispersion in the stained tissue core. **(f)** Normalized histogram of the image in (e) over the foreground (tissue region).

This spectral analysis sheds light on why a simple division by the standard deviation of $\phi(x,y)$ results in the stain-independent map $Z(x,y)$ [Eq. (4) in main text]. The slight difference in central wavelength between incident light and cSLIM (589 nm versus 558 nm) would indicate that cSLIM should have different phase values from gray scale imaging even for *unstained* tissue. However, division of $\phi(x,y)$ by its standard deviation removes the central wavelength dependence of $Z(x,y)$

$$Z(x,y) = \frac{n(x,y) - mean\{n(x,y)\}}{std\{n(x,y)\}}. \tag{S3}$$

Here it is assumed that the *unstained* tissue refractive index $n(x,y)$ is independent of wavelength as well (a reasonable assumption given the small difference in central wavelength between incident light and cSLIM spectra). The operators *mean*{.} and *std*{.} refer to the spatial mean and spatial standard deviation, respectively.

For *stained* tissue, the central wavelength $\lambda'_c$ itself is $x$ and $y$ dependent and does not cancel out when $\phi(x,y)$ is divided by its standard deviation. From Eq. (4) in the main text, the normalized phase image for stained tissue $Z'$, measured by cSLIM, is thus given by

$$Z'[x,y,\lambda'_c(x,y)] = \frac{\dfrac{\Delta n'[x,y,\lambda'_c(x,y)]}{\lambda'_c(x,y)} - mean\left\{\dfrac{\Delta n'[x,y,\lambda'_c(x,y)]}{\lambda'_c(x,y)}\right\}}{std\left\{\dfrac{\Delta n'[x,y,\lambda'_c(x,y)]}{\lambda'_c(x,y)}\right\}} \quad \text{(S4a)}$$

$$Z'[x,y,\lambda'_c(x,y)] = \frac{\dfrac{\Delta n'(x,y,\lambda_c) + \delta n'[x,y,\delta\lambda'_c(x,y)]}{\lambda_c + \delta\lambda'_c(x,y)} - mean\left\{\dfrac{\Delta n'(x,y,\lambda_c) + \delta n'[x,y,\delta\lambda'_c(x,y)]}{\lambda_c + \delta\lambda'_c(x,y)}\right\}}{std\left\{\dfrac{\Delta n'(x,y,\lambda_c) + \delta n'[x,y,\delta\lambda'_c(x,y)]}{\lambda_c + \delta\lambda'_c(x,y)}\right\}} \quad \text{(S4b)}$$

where $\lambda_c$ is the cSLIM central wavelength (558 nm) in the absence of tissue absorbance and $\delta\lambda'_c$ is the shift in central wavelength caused by staining such that $\lambda'_c(x,y) = \lambda_c + \delta\lambda'_c(x,y)$. $\Delta n'[x,y,\lambda'_c(x,y)] = n'[x,y,\lambda'_c(x,y)] - n_0$ with $n_0$ the refractive index of the medium surrounding the tissue and $n'$ the refractive index of *stained* tissue. $\delta n'$ is the shift in stained tissue refractive index due to stain induced dispersion i.e $n'[x,y,\lambda'_c(x,y)] = n'(x,y,\lambda_c) + \delta n'[x,y,\delta\lambda'_c(x,y)]$.

Equations (S4a) and (S4b) would indicate the need for a local normalization constant rather than the global one we have used in our study, in order to make the results from stained and unstained tissue comparable. However, if $\delta\lambda'_c$ and $\delta n'$ are small valued and/or have weak $x$ and $y$ dependence (small spatial variance), Eq. (S4b) becomes approximately

$$Z'(x,y,\lambda_c) = \frac{n'(x,y,\lambda_c) - mean\{n'(x,y,\lambda_c)\}}{std\{n'(x,y,\lambda_c)\}}. \tag{S5}$$

Due to weak dispersion, $n'(x,y,\lambda_c) \approx n(x,y)$, meaning Eq. (S5) is approximately equal to Eq. (S3) and $Z' \approx Z$. As shown in Fig. S5, the $x$ and $y$ dependence of $\delta\lambda'_c$ is indeed small (standard deviation of 1.5 nm over the image). To determine this for $\delta n'$, the following procedure was employed. The 3D spectrum $S'_c(x,y,\lambda)$, was first rescaled and resampled to obtain the frequency spectrum $S'_c(x,y,\omega)$ where $\omega = 2\pi c/\lambda$, $c$ being the speed of light in air. The same was done to $S_c(\lambda)$ to obtain $S_c(\omega)$. The transmission spectrum $T(x,y,\omega)$ was then obtained as

$$T(x,y,\omega) = \frac{S'_c(x,y,\omega)}{S_c(\omega)}. \tag{S6}$$

From the transmission spectrum, the refractive index of stained tissue $n'(x,y,\lambda)$ can be obtained by using the Hilbert transform relationship between the real and imaginary parts of the electric susceptibility [4,5]. The procedure for this is outlined in Ref. [4]. The resulting refractive index is determined only up to an additive constant since the Hilbert transform of a constant is zero. Since we have knowledge of the central wavelength at each pixel [given by $\lambda'_c(x,y)$], the refractive index map $n'[x,y,\lambda_c(x,y)]$ at this central wavelength can be computed from $n'(x,y,\lambda)$. Finally, by subtracting the refractive index map in absence of dispersion, $n'(x,y,\lambda_c)$, from $n'[x,y,\lambda_c(x,y)]$ we get the shift in refractive index due to tissue absorption $\delta n'[x,y,\lambda_c(x,y)]$ or simply $\delta n'(x,y)$. This subtraction also accounts for the differences in additive constants across the $n'[x,y,\lambda_c(x,y)]$ map caused by the computation of an independent Hilbert transform per pixel. $\delta n'(x,y)$ is shown in Fig. S5 (e) whereas Fig. S5 (f) shows its histogram, once again computed only over the foreground (tissue region). As shown, not only is there a small shift in refractive index due to dispersion in tissue but the $x$ and $y$ dependence is also weak (standard deviation of 6.6 x 10$^{-4}$).

Thus, we conclude that our normalization works well despite tissue absorbance because of the small change in $\lambda'_c$, and thus in refractive index $n'$, across the tissue core. Any constant (spatially invariant) changes in both wavelength and refractive index are accounted for in $Z(x,y)$ by subtraction by the mean and division by the standard deviation of the raw phase $\phi(x,y)$.